\pdfoutput=1
\documentclass[%
reprint,
showpacs,
preprintnumbers,
showkeys,
amsmath,amssymb,
prd,
nofootinbib,
]{revtex4-1}

\newcommand{\TITLE}{The Contemporary State of Fundamental Physical Research \\
and the Future Path to Scientific Knowledge}

\RequirePackage[T1]{fontenc}

\usepackage{amssymb,amsmath,latexsym}

\RequirePackage{graphicx}
\RequirePackage{mathptmx}      
\RequirePackage{flushend}
\RequirePackage[colorlinks,citecolor=blue,urlcolor=blue,linkcolor=blue]{hyperref}

\usepackage{sidecap}
\usepackage{ifthen}
\usepackage{keycommand}
\usepackage[toc,page]{appendix}
\usepackage[labelfont={small,bf},textfont=small]{caption}
\usepackage[labelfont={small,bf},textfont=small]{subcaption}

\usepackage{placeins} 

\makeatletter \let\cl@chapter\relax \makeatother 

\usepackage{cleveref}  
\newcommand{\refp}[1]{(\ref{#1})}

\crefname{equation}{eq.}{eqs.}
\crefname{section}{sect.}{sects.}
\crefname{chapter}{chapter}{chapters}
\crefname{table}{table}{tables}
\crefname{figure}{fig.}{figs.}
\crefname{appsec}{appendix}{appendices}
\crefname{appchap}{appendix}{appendices}

\renewcommand\Im{\operatorname{Im}}

\makeatletter
\def\blfootnote{\xdef\@thefnmark{}\@footnotetext}
\makeatother

\graphicspath{{figures/}}

\DeclareCaptionJustification{justified}{\justifying}
\begin{document}
\title{\TITLE}
\author{Milo\v{s} V.~Lokaj\'{i}\v{c}ek}
\email{lokaj@fzu.cz}
\author{Ji\v{r}\'{\i} Proch\'{a}zka}
\email{jiri.prochazka@fzu.cz}
\affiliation{The Czech Academy of Sciences, Institute of Physics}


\begin{abstract}
Classical physics has enabled the acquisition of significant knowledge of the physical properties of nature on a standard macroscopic scale. These achievements were driven by use of the causal ontological approach (proposed originally by Aristotle) to formulate models of physical reality. At the beginning of the 20th century, however, the physics community began to prefer models based on a fundamentally different approach to human knowledge. Copenhagen quantum mechanics (CQM) was used to describe the micro-world. The special theory of relativity was used to describe the kinematics of objects moving at high velocity values in both the macroscopic and microscopic regions. This phenomenological approach to knowledge has been more focused on how things appear - instead of their actual properties and causal sequence. In the middle of the 20th century, the causal ontological approach was used to develop a significant scientific advance: the systematization of fundamental strongly interacting particles on the basis of unified algebra in three-dimensional isotopic spin space for spin values 1, 1/2, and 0. However, it was later strongly deformed under the influence of the phenomenological approach and the quark model. This paper will show that practically all contemporary theoretical models of physical reality contain mistakes or unresolved problems. Further scientific progress can be obtained if (and only if) scientists return to the successful causal ontological approach and falsification. Classical physics may be slightly generalized to enable the description of inertia mass increase in dependence on velocity, replacing the relativity theory and CQM. New assumptions may then be employed using generalized classical physics (GCP) to formulate new descriptions of observed phenomena that were previously inaccurately interpreted and used to promote fundamentally inadequate theories of physical reality.
\end{abstract}


\maketitle

\section{Introduction}
Contemporary scientific and technological progress has been driven by 19th century classical physics and its approach to knowledge. The 19th century was a time when European values, closely associated with scientific knowledge, were accepted by nearly the entire educated world. The advance has been given by using causal ontological approach to knowledge, as proposed in principle by Socrates and developed by Aristotle. Corresponding models of physical reality were compared to observations and further developed, step by step, when there was a contradiction.

In the 20th century, this approach fell into disuse in favor of the phenomenological approach. In the phenomenological approach, it has been sufficient to describe measured quantities only mathematically. More or less meaningless functions have been employed to "understand" a physical system and to make far-reaching conclusions concerning physical reality. There has been little interest in establishing the real properties of these systems and the causes responsible for time evolution. A complete picture of the observed physical system is thus lacking. Systematic analyses of the assumptions (and consequences) behind these theories are also lacking. It can therefore be said that the phenomenological approach has essentially impeded any further actual scientific progress.

Three theories have been applied to physical reality: classical physics in the standard macroscopic region; Copenhagen quantum mechanics (CQM) in the microscopic region; and the special theory of relativity (STR) in the macroscopic and microscopic regions in cases of systems consisting of objects having high velocity values. The transitions between these different regions and theories have, however, remained unexplained. The situation became more complicated when the existence of an inertial mass increase with rising velocity was observed in the first half of the 20th century (with the help of electron accelerators).

Generalized classical physics (GCP), employing a slightly generalized version of Newton's law, can be substituted for all three theories. The time change of a particle impulse (not directly acceleration) is determined by a corresponding force:
\begin{equation}
F = \frac{\text{d}p}{\text{d}t} = \frac{\text{d} \left(m(v) v\right)}{\text{d}t} \, .
\label{eq:force}
\end{equation}
It is necessary only to substitute the constant rest mass values of individual particles in Hamilton's equations with corresponding rising velocity functions. This means that all known observed characteristics of physical reality may be described on the basis of the causal ontological approach. It is possible to apply GCP to all regions of physical reality.

This paper is structured as follows: The contemporary situation in scientific research will be described in \cref{sec:contemporary_situation}. The evolution of scientific knowledge since the end of the Middle Ages will be explained in \cref{sec:evolution}. Changes in scientific approach that were introduced in the past century will be described in \cref{sec:changes}. \Cref{sec:main_problems} will summarize the main problems related to the CQM and the STR. One more important problem concerning classification of fundamental particles will be mentioned in \cref{sec:main_problems}. Other mistaken assumptions in contemporary physical research (some contained in theoretical alternatives commonly used since the 19th century) will be explained in \cref{sec:further_problems}. It will be explained in \cref{sec:future_way} that further scientific and technological progress depends on returning to the successful causal ontological and falsification approaches on which basis classical physics evolved. It will be shown in \cref{sec:future_way} that GCP may be applied to microscopic as well as macroscopic regions at low and high velocity values and therefore enables further progress by providing a theoretical framework for removing mistaken assumptions contained in other widely used contemporary theories.

\section{\label{sec:contemporary_situation}The Current Situation in Science}

Contemporary theoretical models constructed on the basis of the phenomenological approach have been formulated using mathematical functions to describe sets of measured quantities. But little progress has been made in explaining the observed quantities in terms of the real physical properties of the system and the physical causal sequences responsible for transitions from an initial to a final state in space (i.e., time evolution). Insufficiently developed and tested models have been used to "precisely" determine some quantities (or free parameters). The "results" of these models are often misleadingly denoted as measurements or data. The models have an unclear or very indirect relation to the properties of the given physical systems. Usually, a systematic analysis of all involved assumptions in \emph{both} methods used for measurement \emph{and} descriptions of given physical systems has not been performed.

Therefore, it has not been possible to sufficiently study the impact of these assumptions on the overall picture of the physical system represented by the models. In many cases, any partial agreement of a model prediction to experimental data has been regarded as sufficient to regard the model as true and experimentally tested. However, most of these experiments and results have not been reproducible by independent groups of researchers. The phenomenological approach has unavoidably led to completely false conclusions regarding physical reality. This situation is an outgrowth of the rejection of basic Aristotelian logic. Consequently, the conclusions concerning physical reality derived via these models have often been claimed as big discoveries. But they do not have any real value to science.

For example, in the case of proton-proton collisions at high energies, the maximum probability of elastic scattering has been attributed to events at zero impact parameter (i.e., to head-on collisions). However, this is a result of analysis of experimental data and has no analogue in the macroscopic region. It has been based on poorly reasoned, or even mistaken, mathematical relations derived from the corresponding collision models. Protons have been accepted as "transparent" objects during elastic collisions. However, the particle structures themselves have never been explained. Many papers devoted to descriptions of elastic collisions have focused on meaningless quantities. They discuss, for example, the determination of the maximal precision of the meaningless "parameter $\rho$", instead of attempting to explain the full physical picture and to solve known errors and questions contained in the models. Even after decades of development of the models, these approaches have not led to any new reliable and relevant result. Additional detail on descriptions of elastic particle collisions is discussed in \cref{sec:further_problems_es}. Papers that sought to analyze the problem in greater detail, and that were submitted to established journals, were rejected on the basis of misleading or completely wrong (i.e., untrue) argumentation.

Since the end of the 20th century, leading journals have refused to publish papers containing ideas on fundamental physics that do not conform to "generally accepted" concepts - even if these concepts are based on assumptions that have been shown to be faulty using a causal ontological approach. Grants and large financial sums have been given to authors who have published in well-known and established journals. They continue to be rewarded based on the number of citations of their papers, even if their work has been proven to be wrong (i.e., falsified) in papers published in less influential journals or in publicly available internet archives (e.g., arXiv.org). In other words, scientific truth has been judged by voting. Truth has been, and continues to be, misconstrued and emptied of meaning.

The current situation was described in a 2013 editorial in \emph{The Economist} entitled "How science goes wrong" \cite{Economist2013}. The article mostly examined biotechnology. But it is shocking to consider that science is currently "going wrong" in practically all regions of fundamental scientific research. There has been no significant reaction from scientists to this situation.

\section{\label{sec:evolution}The Evolution of Scientific Knowledge in the Modern Period}

To comprehend the contemporary situation, it is necessary to examine the evolution of scientific knowledge. It can be said that the development of classical physics represents the greatest scientific progress in human history. The achievements of G.~Galilei (1564-1642) and I.~Newton (1643-1728), regarded as the founders of classical physics, were brought about by their use of the causal ontological approach, which was originally proposed by Socrates (470-399 BC) and developed by Aristotle (384-322 BC). This approach was not accepted in Europe until the 13th century, when it was embraced by Albert the Great (1200-1280) and other scholars, who learned of it from Islamic scientists. Texts credited to Avicenna (980-1037) and Averroes (1126-1198), which expanded on Ancient Greek philosophy, were intensively studied in Europe. Thomas Aquinas (1225-1274) further developed the causal ontological approach. It was based on the assumption that the world in which we live represents the fundamental source of our rational knowledge. Causal relations (assumptions) between different states and the main parts of physical systems were formulated based on observations of the time evolution of corresponding processes. Consequences arising from these assumptions were deduced and tested. If agreement with corresponding experimental data was obtained, then the assumptions were accepted as plausible and the consequences were developed further. When a contradiction was discovered, it also represented important knowledge. A contradiction led to the necessity of rejecting the incorrect set of assumptions and the obligation of finding new set of assumptions that would correspond to observations.

The falsification approach originally applied to the physical region (i.e., non-living systems) and was later applied to individual living entities (i.e., organisms). Many properties of organisms cannot be explained on purely physical grounds. Different types of organisms may exist on different levels of complexity. The properties of lower-level organisms are not adequate to fully describe all properties of higher-level organisms (e.g., to explain human beings and human society). Additional factors must be included to explain the transition from a lower level to a more advanced level. Without an additional cause, human beings could not have evolved from lower-level organisms. 

The formulation of human ethical rules was strongly influenced by the falsification approach. In the 19th century, so-called European values, formulated with the help of the causal ontological approach and falsification, were accepted by almost the entire educated world. All contemporary technological progress has been based on this approach to knowledge. Yet in the modern age, this successful approach has been substituted by use of the insufficient phenomenological approach, not only in technological areas but also in areas such as human ethics, morality and law.

Different possible principles of human knowledge began to be discussed around the end of the Middle Ages. Some philosophers wanted to expand interpretations of the so-called "first cause", and a path parallel to the classical route developed. R.~Descartes (1596-1650) attempted to solve the problems. However, his formulation "Cogito, ergo sum" ("I think, therefore I am") most likely had a different (i.e., nearly opposite) meaning for him than the interpretation formulated later on the basis of a purely phenomenological approach (see J.~Locke (1632-1704), D.~Hume (1711-1776) and A.~Comte (1798-1857)). The parallel path to the classical route developed in a positivistic way, based on a phenomenological approach to knowledge. Many philosophers overestimated the potential of human reason - and underestimated the importance of testing and the falsification approach.

The phenomenological approach made a significant impact on physics in 1867, when L.~Boltzmann (1844-1906) declared as a physical law the increase of entropy (i.e., the phenomenological characteristic of a macroscopic system was accepted without considering the real physical processes and properties of the system). Other bigger, fundamental shifts in scientific thinking followed in the 20th century, resulting in a series of mistaken statements that were denoted as authentic physical results.

\section{\label{sec:changes}Changes of Scientific Thinking in the 20th Century}
The earlier mistaking trend was supported significantly by N.~Bohr (in 1913) when basic physical law concerning atom structure was represented by two phenomenological postulates (one concerning the existence of individual energy levels and the other one trying to bring the differences between these levels to harmony with the measured spectra of emitted photons). These two postulates have been taken as basic natural laws without looking for any other reasons of corresponding matter structure. The contemporary situation has been caused, of course, mainly by the further step of Bohr (in 1927) when he added some quite arbitrarily chosen (simplifying) assumptions to Schr\"{o}dinger equation and formed Copenhagen quantum mechanics (CQM) \cite{Bohr1928}. He deformed fully the physical interpretation of the solutions of Schr\"{o}dinger equation that were interpreted originally by Schr\"{o}dinger in the standard classical sense (particle interpretation) \cite{Schroedinger1926}. Common conviction has existed that Schr\"{o}dinger equation has differed from classical physics even without this deformation by Bohr. It has been, however, shown in the end of the 20th century that this equation may be derived in the framework of classical physics which will be discussed in \cref{sec:main_problems_sch_eq}.

As to the CQM it was criticized by Einstein on the basis of causal ontology. He showed (with two coauthors B.~Podolsky and N.~Rosen) \cite{Einstein1935} in 1935 with the help of a Gedankenexperiment (denoted commonly as the EPR experiment) that the theory required the existence of immediate interaction between two very distant matter objects. He assumed that two particles produced by a particle decaying in rest and moving in opposite directions may be detected by two detectors:
\begin{equation}
|| <---- \; 0 \;  ----> ||
\label{eq:polarizers}
\end{equation}
He demonstrated that in the theory the result established in one detector determined immediately, too, what happened in the other detector, or that both the particles had to be in mutual link (denoted as entanglement at the present), which has been fully unacceptable in causal ontology. Bohr refused Einstein's criticism by arguing that it was based on observations in macroscopic world while in the microscopic region the properties may exist. The physical scientific community supported strongly Bohr's standpoint while Einstein did not abandon his standpoint till the end of his life. Since that time two quite different theories have been assumed to be valid in macroscopic and microscopic regions of matter world, even if nobody has been trying to find a boundary between them or to study how macroscopic objects may consist of so strange microscopic particles.

The alternative hidden-variable theory appeared then in 1952 when D.~Bohm \cite{Bohm1952} discovered that an additional ("hidden") variable was contained in the solutions of Schr\"{o}dinger differential equation; the CQM could not be valid in such a case. And the differences between these two quantum theory alternatives started to be studied.
Later it was shown that also the causal interpretation of the equation proposed in principle originally by Schr\"{o}dinger was to be admitted to hold for microscopic particles (as only spiral solutions have been excluded), which will be discussed in more details in \cref{sec:main_problems_sch_eq}.

J.~Bell \cite{Bell1964} tried to decide between these two alternatives. He generalized the EPR Gedankenexperiment having assumed that both the particles running in opposite directions had also opposite spins (two polarizers were assumed to be added):
\begin{equation}
|| <---- |^{\alpha}----\; 0 \;  ---- |^{\beta}----> ||
\label{eq:polarizers_spin}
\end{equation}
where $\alpha$ and $\beta$ denoted angle orientations of the corresponding polarizers.
The coincidence probabilities of two particles having passed through differently oriented polarizers were then measured. Having assumed the existence of a limiting condition between some individual passage probabilities Bell derived (in 1964) the following condition for the special combination of four passage probabilities at different polarizer angles:
\begin{equation}
B = a_{\alpha_1} b_{\beta_1} + a_{\alpha_1} b_{\beta_2}+a_{\alpha_2} b_{\beta_1} - a_{\alpha_2} b_{\beta_2} \leq 2
\label{eq:bell_ineq}
\end{equation}
It has been then commonly assumed (without any test having been performed) that the condition has been valid in Schr\"{o}dinger's particle alternative (not contradicting Einstein's criticism) but invalid in Bohr's one. The corresponding experiment was proposed and performed in 1982 \cite{Aspect1982} and it has been concluded
\begin{itemize}
\item{Bell's inequalities were clearly violated at some angle combinations;}
\item{probability of passage of two photons running in opposite directions through polarizers (as shown in \refp{eq:polarizers_spin}) in dependence on angle between transmission axes of the polarizers was expressed in the form of the Malus's law characterizing the passing of unpolarized light through two polarizers in standard experiments measuring transmission in dependence on angle between their transmission axes \cite{Aspect2004}.
}
\end{itemize}
Since that time the CQM has been regarded as the only theory of microscopic world, even if it has been demonstrated later that the inequality \refp{eq:bell_ineq} may hold in a classical deterministic case only.

The attention to the problems concerning consequences of the CQM has been later called also in the book of R.~Newton \cite{Newton2009}, having the subtitle: "Einstein was correct, but Bohr won the game". However, any new results have not been brought and Bohr's victory has been taken commonly as the final fact. The problems have been studied, however, very intensively in Prague in the last 20 years of the past century and new results have been obtained \cite{Lokajicek2012_controversy,Lokajicek2013,Lokajicek2012_bell,Lokajicek2014}. It has been possible to show that the preference of Bohr's theory has been based on two mistaking assumptions:

\begin{itemize}
\item{it has been assumed all the time that the additional assumption involved in derivation of Bell's inequalities has corresponded to the hidden-variable theory of D.~Bohm, while in fact it has corresponded to basic classical physics only (where no spins of photons have been taken into account);}
\item{it has been stated without any reasoning that the hidden-variable theory has had to give the predictions differing significantly from the Malus's law.}
\end{itemize}

It means that any argument does not exist against Einstein's conclusion based on causal ontological approach. Einstein was evidently right; and in the past century everyone was forced to believe in quite unphysical characteristics of the matter world. The remaining problem of Schr\"{o}dinger equation and its interpretation will be further discussed in \cref{sec:main_problems_sch_eq}.

\section{\label{sec:main_problems}The Main Contemporary Open Theoretical Problems}

It is widely accepted that classical physics is valid in the standard macroscopic region, while quantum physics, based on Schr\"{o}dinger's equation, should be applied in the microscopic region. However, results differing from reality have been obtained in many cases of CQM proposed by Bohr. In both the macroscopic and microscopic regions, the special theory of relativity (differing fundamentally from classical physics and CQM) has been applied at high velocity values.

Problems and questions concerning the interpretation of Schr\"{o}dinger's equation (which strongly concerns the foundations of CQM) and inertia mass increase with velocity (which is closely tied to the dependence derived in the framework of STR) are described in the next two subsections. One further important problem concerning the classification of strongly interacting particles (the deformation introduced by SU(3) group) is discussed in the third subsection.

\subsection{\label{sec:main_problems_sch_eq}Schr\"{o}dinger's Equation and its Interpretation}

It is widely believed that Schr\"{o}dinger's equation differs from classical mechanics. However, as mentioned previously, it was shown at the end of the 20th century that solutions of Schr\"{o}dinger's equation represent set of superpositions of Hamilton's equations solutions if spiral solutions (existing in the case of attractive force values) have been excluded (see \cite{Ioannidou1982,Hoyer2002,Lokajicek2007_schroedinger_eq,Lokajicek2012}). In particle interpretation, Schr\"{o}dinger equation solutions have always been represented by linear combinations (superpositions) of simple solutions of classical Hamilton's equations, represented by corresponding basic states (according to the corresponding initial condition). Consequently, Schr\"{o}dinger's equation is valid in both the macroscopic and microscopic regions, at least for small velocity values in agreement with Hamilton's equations.

The advantage of Schr\"{o}dinger's equation has existed mainly in the possibility of describing directly the results of the evolution of a system if its initial state has been represented by the superposition of a set of basic states represented by individual solutions of Hamilton's equations. Solutions to Schr\"{o}dinger's equation must, of course, be interpreted on the basis of particle description as it was proposed originally by Schr\"{o}dinger in 1926 \cite{Schroedinger1926}). However, this was completely excluded by Bohr in 1928 (see details in \cref{sec:changes}). Bohr insisted that incoming and outgoing states (represented by the same coordinate dependence of amplitude and differing by impulse sign only) are represented by one vector in the corresponding Hilbert space, while they lie in two mutually orthogonal subspaces (each involving whole coordinate space) \cite{Lax1976}. More details related to the interpretation of Schr\"{o}dinger's equation and CQM may be found in \cite{Lokajicek2007_physteor,Lokajicek2012_controversy,Lokajicek2013,Lokajicek2012_bell}.

\subsection{\label{sec:main_problems_inertia} Inertia Increase with Velocity and the Theory of Relativity}

The special theory of relativity differs from classical mechanics and CQM mainly at high velocity values. However, even after more than a century of its existence \cite{Einstein1905}, there are aspects of the STR that have never been sufficiently explained and are still questioned and doubted (e.g., see \cite{Jacobs2009}, a detailed treatise that includes historical context and many critical comments). Several basic assumptions (or consequences) of the theory are often regarded as counterintuitive. For example:

\begin{enumerate}
\item{\emph{finite maximum velocity}\\
The velocity of any moving object cannot exceed the speed of light in a vacuum $c$. The relative velocity of two particles is allowed only to be lower or equal to $c$ (the relative velocity of two particles, each having speed $c$ or very close to this value, and moving in opposite directions is, however, determined to be $c$ and not $2c$, as one would expect).
}
\item{\emph{invariance of the speed of light}\\
Light (a moving photon) in a vacuum has the same value of velocity $c$ in any inertial reference frame. Taking into account that the speed of any particle or object in a rest frame is zero by definition, it is "not allowed" to use a photon rest frame in this theory (the photon rest frame is inapplicable for a description of physical reality in this theory). Invariance of the speed of light represents a fundamental assumption of the entire relativity theory \cite{Einstein1905}. It is often argued that it is an experimental evidence, on the basis the famous Michelson-Morley experiment of 1887 \cite{Michelson1887}. However, interpretation of the experiment is complicated by many assumptions. A detailed discussion of the experiment, showing that the problem is more complicated than is usually presented (and including the historical context of motivations of many authors working in this field at that time) is found in \cite{Jacobs2009}.
}
\item{\emph{zero rest mass of photon}\\
It is assumed that a photon has zero rest mass and, therefore, may move by the speed $c$ in a vacuum. However, in some physical applications, the non-zero rest mass had to be assumed if certain calculations were to be performed. In such cases, the rest mass has been put at zero value (if the corresponding limit has been calculated) only in final formulae.
}
\item{\emph{relative time (time dilation)}\\
The time elapsed in one reference frame is different from that elapsed in another inertial reference frame if it is moving with respect to the first reference frame.
}
\item{\emph{relative length (length contraction)}\\
Similarly, the length of objects is relative, in dependence on their speed with respect to a given reference frame.
}
\item{\emph{relativity of simultaneity}\\
Two simultaneous events at different places in one reference frame may not be simultaneous in another reference frame (in dependence on the choice of the given reference frames).
}
\item{\emph{invariance of spacetime interval}\\
Spacetime interval between any two events is the same in any inertial frame. The time coordinate is correlated to space coordinates in dependence on speed of light $c$ in a vacuum.
}
\end{enumerate}

We point out another problem that is not widely known: In the framework of the STR, it has been derived that inertial mass will increase with velocity. It has been derived that the kinetic energy of any particle increases with velocity. It is to hold $E_{\text{tot}}(v) = E_0 + E_{\text{kin}}(v)$ where
\begin{equation}
E_{\text{kin}}(v) = m_0 c^2 \left(\gamma(v) -1 \right)
\end{equation}
and
\begin{equation}
\gamma(v) = \frac{1}{\sqrt{1-\left(\frac{v}{c}\right)^2}}
\end{equation}
is the Lorenz gamma factor. $E_0 = m_0c^2$ represents additional rest energy derived for inertial mass at zero velocity. It has been concluded from the gamma factor that the velocity of any particle cannot exceed the speed of light. It has been derived that the inertial mass is to increase with rising velocity:
\begin{equation}
m(v) = m_0 \gamma(v) \; .
\end{equation}

At low velocity values, the relativistic formula for kinetic energy does not differ significantly from the classical formula $E_{\text{kin}}(v)= \frac{1}{2}mv^2$ (where $m$ has been regarded as constant). Very great differences exist at higher velocity values.

The validity of these formulae concerning inertial mass increase with velocity has been fully accepted by the physics community. This occurred after some increase of inertial mass with increasing velocity was found in the first half of the 20th century (with the help of electron accelerator data). The agreement of the predicted and measured dependences has been commonly regarded as completely sufficient. However, comparison of these very basic relativistic formulae to corresponding experimental data has been very limited. The systematic measurement of the inertia mass increase of different particle types - across a very broad range of velocities (including close to the speed of light) - should be performed and analyzed independently of the assumptions of the STR.

As mentioned in the introduction, Newton's classical mechanics (corresponding to constant mass) may be generalized if force (acting on a particle) is assumed to define the time change of particle momentum (see \cref{eq:force}), instead of particle acceleration according to $F=m a$, where $m=m_0$ is independent of velocity. The increase of inertial mass may be derived in the framework of standard Hamilton's equations. In GCP, it is necessary only to substitute the constant rest mass values of particles by corresponding velocity functions in the definition of Hamiltonian in Hamilton's equations.

In the GCP framework, the various velocity dependencies of inertial masses may be derived at high velocity values (in dependence on the values of existing free parameters). In special cases, dependence, corresponding to Newton's classical physics or to the STR, may also be obtained \cite{Lokajicek2010}. Detailed theoretical predictions may be derived for corresponding experimental arrangements and free parameters determined on the basis of experimental data.

\subsection{\label{sec:main_problems_particles} Particle Systematics: Original Isotopic Spin Algebra and the SU(3) Group}

The systematics of fundamental strongly interacting particles (hadrons) was proposed on the basis of isotopic spin symmetry, when different electric charges of these particles have been characterized using three-dimensional isotopic spin space (instead of four-dimensional) \cite{Votruba1958_isospin,Votruba1958_algebra,Lokajicek1958_algebra}. In 1958, a summary of this theory, using an algebra for the classification of hadrons, was presented at the Rochester conference at CERN in Geneva \cite{Espagnat1958}. In the algebra, descriptions of particles having half and unit spins were joined. The algebra has eight basic elements that are to be applied to each of three kinds of basic strongly interacting fundamental particles: baryons, antibaryons and mesons (existing antimesons are included in the last octet). All these particles decay relatively slowly (with the exception of protons and antiprotons that may be taken as stable in a vacuum).

It may be assumed (based on experiments and the causal ontological approach) that each hadron is a dynamic system consisting of different subsystems which are bonded by standard strong interaction. These subsystems may be formed by a great number of some super strongly bonded elementary particles. If two subsystems have been formed, and a subsystem obtains sufficiently strong deviating momentum, the given hadron may decay and the subsystems become separate objects (which may or may not further decay).

In higher energy collisions of hadrons, some excited states may be formed (having much greater rest mass values). They should correspond to the products of greater numbers of basic octet algebra elements. These states, as well as corresponding subsystems, should be in agreement with this requirement. This holds also for states arising in consequent decays.

It is clear that, essentially, only the formation of two subsystems may lead to corresponding decay particle processes. Different subsystem pairs can probably be formed in all hadrons. These pairs may be changeable. The pattern of a decaying hadron may be substantially influenced by the subsystem pair. Different numbers and types of decaying hadrons may, for example, be created in proton-proton collisions. The result may be very different in dependence on impact parameter values and geometrical orientation of two subsystem pairs existing in the two protons at the moment of collision. The corresponding subsystem pairs may also be responsible for the creation of a different number of jets of different sizes, in dependence on collision energy values.

The (weak) decay characteristics of the basic octet particles were analyzed immediately after the presentation of the results in the middle of the 20th century. Much more experimental data is now available concerning the formation of many new objects produced in, e.g, proton-proton collisions. It allows testing of various aspects concerning the structure and systematization of hadrons in greater details than it was possible in the past. However, a detailed analysis can only be done on the basis of the causal ontological approach. No new theoretical scientific results (insight into the physical processes) can be obtained using the phenomenological approach.

The analysis of particle properties and their systematization was deeply influenced by political conditions in Europe in the 1950s and 1960s. The evolution of corresponding ideas occurred without the participation of the original authors. The approach introduced in \cite{Votruba1958_isospin,Votruba1958_algebra,Lokajicek1958_algebra} was described as an \mbox{"eightfold way"}, but made no mention of the original papers. It was modified (deformed) when the original isotopic spin algebra was replaced by the SU(3) group. Baryons were assumed to correspond to 3 quarks and mesons to quark-antiquark pair. However, much more complicated combinations of quark or anti-quarks were attributed to some hadrons, even from basic multiplets (e.g., to $\pi^0$). Quarks have not yet been observed. The systematics of hadrons based on the SU(3) group differs significantly from the originally proposed systematics based on the isotopic spin algebra. Some quite arbitrary assumptions were added in the SU(3) group approach. Models of hadron collisions based on this approach cannot help in understanding the pattern of experimental data established at different energy collision processes. Consequently, the contemporary classification of hadrons based on the SU(3) group approach, and the interpretation of corresponding quarks as constituent particles of hadrons, is hardly acceptable.

\section{\label{sec:further_problems}More Mistaken Assumptions}

The preceding section concerned itself with conceptual problems. Mistaken assumptions, however, have also been included in models of individual physics problems. These assumptions should be analyzed and resolved if our knowledge of fundamental reality is to expand.

\subsection {Collision S-Matrix and its Unitarity}

Collision processes have usually been described with the help of an S-matrix (to determine the transition probabilities from individual incoming states to outgoing states). In contemporary approaches, as required in CQM, all possible elastic states of two colliding particles have been assumed to be represented by vectors in very simple Hilbert space (without any subspaces). The corresponding vectors in Hilbert space have been defined on the basis of corresponding wave functions only. Its derivatives, according to space coordinates (direction of corresponding momentum), have not been taken into account. It has been required for this S-matrix to be unitary, which has been interpreted as the consequence of probability conservation.

Such a requirement, however, is not acceptable in Hilbert space, where incoming and outgoing states of colliding particles belong to different subspaces. Other outgoing subspaces must be added if some inelastic collisions also exist (see \cref{sec:main_problems_sch_eq}). In such cases, the S-matrix cannot be unitary - the sum of elements expressing the transition probability from one incoming state to the set of all possible outgoing states should be equal to one. This problem is analyzed in greater detail in \cite{Prochazka2015}.

\subsection{\label{sec:further_problems_es}Elastic Scattering, the Optical Theorem, and Characteristics of Incoming Collision States}

In the case of the elastic scattering of two particles, it is commonly assumed that the so-called optical theorem is valid:
\begin{equation}
\sigma_{\text{tot}} \sim \Im F(\theta\!=\!0) \; ;
\end{equation}
That is, the imaginary part of scattering amplitude $F$ at zero scattering angle $\theta$ is required to be proportional to the total cross-section. The theorem was used in optical analyses on the basis of experimental results. However, its validity has never been determined exactly in particle physics \cite{Prochazka2015}. The influence of the distribution of incoming states, in dependence on impact parameter $b$, has not been considered in attempts to derive an optical theorem in particle physics. Great deviations from reality may be obtained if this method is used in strong interactions, when only a very small part of incoming states is influenced by these interactions.

According to nearly all contemporary models of elastic collisions of (charged) hadrons, elastic collisions should be interpreted as central processes: The mean impact parameter corresponding to elastic collisions should be lower than that of inelastic collisions. The structure of colliding particles that should correspond to this behavior has, however, not been explained in the literature. This kind of transparency of particles during elastic collisions, even at $b=0$ (i.e., head-on collisions), does not correspond to the usual ideas of collisions of two matter objects. It was shown as early as 1981 that the models suggesting centrality of elastic collisions were based on arbitrarily and unreasonably chosen assumptions.

Mathematical models may be, however, modified to obtain the peripheral behavior of elastic scattering at high energies. It has been shown explicitly that protons may be interpreted as rather compact (non-transparent) particles that may collide elastically only at higher values of impact parameter. They break up at high collision energy values if the corresponding value of impact parameter is very low. A detailed discussion of analyses of experimental data under different assumptions (using the so-called eikonal model) and their impact on characteristics of protons may be found in a recent paper \cite{Prochazka2016_eikonal} that also includes a summary of historical context.

There are other unresolved problems with contemporary descriptions of the (elastic) scattering of two particles (see sect.~6 in \cite{Prochazka2016_dependence}). In \cite{Prochazka2016_dependence}, one can find an historical context for descriptions of elastic collisions of (charged) hadrons in dependence on impact parameter. It is necessary to start from a causal ontological approach and to introduce the probabilities of collisions in dependence on impact parameter to solve the problems and open questions. In sect.~6 in \cite{Lokajicek2013}, one may find a very preliminary model of elastic particle collisions that corresponds to ontological approach requirements and that has been applied to experimental data. This new probabilistic model explicitly shows how to describe elastic scattering as a peripheral collision process, without using the optical theorem. It also shows how to solve some problems discussed in \cite{Prochazka2016_dependence}. One expects it will be possible to obtain even better results if the model is further developed and outstanding problems addressed.

\subsection{Different Kinds of Force Actions}

In the framework of standard classical physics, only forces that correspond to electromagnetic and gravitation potentials have been considered. These forces are non-zero at any distance and diminish with increasing distance. This is usually described with the help of a potential whose source is in point centers (represented by individual particles in a closed system). However, in the microscopic region, strong forces also exist that can hardly be described by a similar potential.

It is clear that the influence of strong interaction should not be interpreted as existing at a greater distance: Strong interactions can be interpreted as contact interactions. But complications arise. These microscopic objects should not be described as spherical. They may, for example, be interpreted more generally as elliptical, with variable lengths of axes corresponding to change caused by internal dynamics. Therefore, a new way of describing the influence of corresponding forces must be found. The comparison of collision results at different impact parameter values, and at different energies, might provide the necessary information. Detailed characteristics derived from analysis of elastic collisions of different particles, with the help of the new probabilistic model (see the end of \cref{sec:further_problems_es}), may be very helpful in this case.

\subsection{Maxwell Equations and Speed of Light}

Another important problem concerns Maxwell equations. They represented decisive progress in physical research on unified electric and magnetic phenomena. However, identification of the velocity of electromagnetic signal transfer with light velocity probably halted further significant progress in this research. As explained in \cite{Krumm1986}, Maxwell obtained for this velocity a very high value, but less than light velocity. He identified the velocities and made them equal (at that time, light velocity was regarded as a fundamental quantity). This conviction holds until now, even if Einstein's discovery of the existence of photons fundamentally changed the previous basic physical concept. It is possible to say that two very diverse physical concepts in this region have been accepted in different cases.

There are, consequently, two fully unresolved questions. The first concerns the basic properties of photons: Their inertial and rest mass values have remained essentially unresolved. The possibility of interpreting photons as a kind of mass object has not been sufficiently tested. The second concerns the problem of electromagnetic signal transfer in open space that might be mediated by some not-yet-identified objects existing in it. How this transfer is related to photons must be analyzed in greater detail. It may not correspond directly to light velocity.

\subsection{General Theory of Relativity and Speed of Gravity}

In 1916, Einstein published \cite{Einstein1916} the general theory of relativity (GTR), which tried to generalize STR and Newton's theory of gravity. GTR is reduced to STR in a limit case of inertial reference frames and no presence of gravity. The problems contained in STR, and discussed in \cref{sec:main_problems_inertia}, are involved and also concern GTR.

It is often argued that classical physics failed to explain the perihelion precession of Mercury and that a formula derived by Einstein in 1915 \cite{Einstein1915_merkury}, in the GTR, explained the effect. Addressing this problem is complicated - and depends on how one chooses to define "classical physics". Newton's classical theory of gravity assumed an infinite speed of gravity and did not predict any shift of Mercury's orbit (which is relatively very small; it took several centuries to observe it). The speed of gravity is assumed to be equal to the speed of light in GTR. It is not widely known that the formula presented by Einstein in 1915 was, in fact, derived much earlier by Paul Gerber, in 1898 (i.e., 17 years before Einstein), in the framework of classical physics \cite{Gerber1898}. Gerber considered the finite speed of gravity and derived, on the basis of observations of the shift of Mercury's orbit, that it should have a value very similar to the speed of light. The question of the explanation of the observed shift of Mercury's orbit is even more delicate due to the fact that several effects are used to explain the total observed shift. The "Gerber-Einstein" formula is assumed to account only for small part of it. Further valuable comments may be found in \cite{Capria1999}.

The value of the finite speed of gravity is an open question. One should acknowledge it may be different from the speed of light. As explained in \cref{sec:main_problems_inertia}, GCP can be used to describe the different dependences of inertia mass increase with velocity. One may, therefore, ask how significant these two effects - the finite speed of gravity and inertia mass increase - are in explaining the perihelion precession of Mercury. Corresponding analyses may be performed in GCP - while STR and GTR are not useful in this case. GCP may also aid the understanding of several other assumptions, such as the existence and origin of dark matter and energy, that have been used to interpret (alongside the GTR) a range of astrophysical observations.

\subsection{Goedel Theorem}

Another significant influence on the evolution of physical thinking in the 20th century was the work published by K.~Goedel in 1931. His famous incompleteness theorems have been commonly interpreted as general proof that a complete and consistent set of assumptions for all mathematics and physics does not exist. This view has corresponded to phenomenological trends and, like CQM, has been generally accepted. However, there has been no known effort to test Goedel's overall conclusions, which continue to exist mainly in the German language.

M.~Hirzel has attempted to translate at least some of the main parts of Goedel's original arguments into English. He has called attention to some significant difficulties: Goedel's original approaches were formulated in an unusual notation. Consequently, Hirzel has declined to express his clear standpoint on Goedel's results (his view may be found in \cite{Hirzel2000}). It is evident that Goedel's theorem can hardly be regarded as an argument acceptable for physical theories.

\section{\label{sec:future_way}The Path to True Scientific Knowledge}
\subsection{\label{sec:future_way_ontology}The Causal Ontological Approach and the Falsification Approach}

As previously mentioned, classical physics represented big step forward in understanding the physical processes around us. Because it was based on the causal ontological approach to knowledge proposed in principle by Socrates and developed by Aristotle, it also led to significant technological progress. In Europe, this approach to knowledge was developed further by Thomas Aquinas on the basis of natural science and a Christian approach to the world. Detailed analysis by K.~Popper (1902-1994) in 1934 \cite{Popper1934} showed that the approach has been based fundamentally on falsification.

The world in which we live represents the basic source of our knowledge and the basis of our reason. One may observe physical phenomena or perform dedicated experiments to better understand it. To explain an observed phenomenon concerning a physical system one may try to formulate, on the basis of logical induction, a theoretical model using a causal ontological approach. That is, one may try to explain the phenomenon in terms of assumed properties of the objects in the system and their mutual interactions, taking into account causal sequences (i.e., causes responsible for time evolution of the system in space). However, any such model of physical reality is always based on assumptions that may or may not be true. In the falsification approach, one should attempt to derive all possible consequences of the employed assumptions. The existence of any contradiction (in the model or in a model prediction when compared to physical reality) must be interpreted as the invalidation of the employed set of assumptions and must be regarded as part of our knowledge. A new set of assumptions used for description of the physical system must be formulated (or at least some of the previous assumptions modified) and tested again. If no contradiction is found, it does not mean that the given assumptions (and all their consequences) are true - it means only that they are plausible. One cannot exclude the possibility that a contradiction may be found later. Our knowledge of the world we live in was derived, step by step, on the basis of the iterative development of different models (i.e., the systematic study of their assumptions) and their comparison to physical reality.

The approach to human knowledge described above and based on causal ontology and falsification represents the only possibility of obtaining accurate knowledge and making further technological progress. However, this approach was rejected as the phenomenological approach described in \cref{sec:contemporary_situation} became dominant. Further details concerning possibilities of true human knowledge may be found in \cite{Lokajicek2007_models}.

\subsection{\label{sec:future_way_gcp}Generalized Classical Physics (GCP)}

All valuable contemporary scientific results have been derived from the classical foundation. The relativity theory and CQM (discussed in \cref{sec:main_problems}), based on the phenomenological approach, were proposed and used to interpret experiments that "could not be explained classically". However, it is possible to generalize classical physics and develop models based on new assumptions to interpret these experiments without resorting to new theories based on the phenomenological approach. One may (and should) interpret these experiments with the help of causal ontological models - something that was not sufficiently done in the past.  

Regarding the dynamics of moving bodies, the main results of classical mechanics have been based on Hamilton's equations, in which Newton's simplifying condition (the constancy of inertial mass at any velocity value) was applied. However, Hamiltonian mechanics may also hold in more general cases, including inertial mass increase with velocity. It is only necessary to substitute the constant mass values of individual particles in the corresponding Hamiltonian in Hamilton's equations by those rising in dependence on velocity (see also \cref{eq:force}). It should then be possible to determine the inertial mass increase with velocity for a given particle type on the basis of experimental data (see the end of \cref{sec:main_problems_inertia}).

It was mentioned in \cref{sec:main_problems_sch_eq} that Schr\"{o}dinger's equation may be derived (under certain limiting conditions) on the basis of Hamilton's equations (when it is rightly interpreted on particle basis according to Schr\"{o}dinger's original proposal). This means that Hamilton's equations may be used for descriptions in both the standard macroscopic and microscopic regions.

GCP (based on generalized Hamiltonian mechanics) may, therefore, be used for descriptions of all matter reality - in the microscopic and macroscopic regions and at any value of velocity. All three theories (classical mechanics, CQM and STR) that are now applied to different regions of reality may be replaced by GCP.

Regarding the classification of fundamental particles, it must be concluded that instead of the contemporary classification based on the SU(3) group and quark interpretation, one should return to the original isotopic spin algebra (see \cref{sec:main_problems_particles}) - that is, to the three-octet system of stable or relatively slowly decaying hadrons. To better understand the structures and interactions of particles (including the classification of particles), greater emphasis should be put on detailed analyses of decays and mutual elastic collisions of individual, relatively slowly decaying and strongly interacting particles (belonging to the three octets). Focus should also be applied to inelastic collisions that produce individual resonances and small numbers of secondary particles. More attention should be devoted, too, to the structures and interactions of small atom nuclei consisting only of several protons and neutrons.

Important information concerning the structure and interaction of colliding particles can be obtained if collision probabilities (which may depend on corresponding impact parameters at different energies) are properly taken into account. However, no contemporary model based on CQM or the theory of relativity has so far been able to provide corresponding $b$-dependent probabilities. A proposed preliminary probabilistic collision model has been mentioned in \cref{sec:further_problems_es}. This model has shown it is possible to analyze (elastic) pp scattering data to obtain interesting characteristics of protons that have not been determined previously. Important information concerning the properties of particles from the basic octets may be obtained if the model is generalized and experimental data of different particle types are analyzed with its help.

The only way to progress in physics, therefore, consists in using GCP. The falsification approach should be used to develop models of physical systems formulated on the basis of causal ontology. This is the only approach that may result in real insights into a physical system.

\section*{Conclusion}
Over the last century, the influence of the phenomenological approach has completely overwhelmed the causal ontological and falsification approaches. The goal of current physics theory consists of formulating purely mathematical models to describe a set of measured values. Usually, none of the assumptions of these models, or the methods used to evaluate the results of corresponding experiments, are examined or even mentioned. Instead, any partial test or comparison to experimental data is considered sufficient to denote these models as true (i.e., fully corresponding to reality). The phenomenological approach, used for descriptions of physical systems in natural sciences (see \cref{sec:contemporary_situation}), has led only to confusion and complications. Over time, (see \cref{sec:evolution,sec:changes}), scientific progress has been slowed or even completely blocked by it. Phenomenological models have always led to unfortunate deficits of insight.

It has been shown that two widely used fundamental theories, CQM and the theory of relativity, are based on doubtful, insufficiently tested or even mistaken assumptions. Use of these theories, whose development was significantly influenced by the phenomenological approach, has impeded real scientific progress. The systematics of strongly interacting particles, based on quarks having strange properties, is another proposition that has negatively impacted real scientific progress (see \cref{sec:changes}). Other mistaken assumptions used in contemporary theoretical descriptions are discussed in \cref{sec:further_problems}.

In this situation, it is clear that new scientific knowledge will be achieved only by returning to the basic approach to knowledge - that is, to causal ontology and falsification (see \cref{sec:future_way_ontology}). All historic and contemporary technological progress has been based on this approach to knowledge. 

In the middle of the 20th century the classification of strongly interacting particles was proposed on the basis of isotopic spin algebra using the causal ontological approach. However, this systematization of the particles was later modified and deformed under the influence of the phenomenological approach, see \cref{sec:main_problems_particles}. To remove these deformations one should return to the original proposal on the basis of isotopic spin algebra and analyze further the classification of hadrons using also large amount of experimental data which were not available in the past.

Important experimental results of the last century have concerned the increase of inertial mass with velocity. This was affirmed in principle on accelerators in the first half of the 20th century. The dependence obtained in the framework of the theory of relativity has been accepted by the physical community, even if corresponding systematic and detailed comparisons to experimental data have never been established (especially at velocities very close to the speed of light). The increase of inertial mass with velocity can be described with the help of Hamilton's equations, where Hamiltonian mechanics has been generalized (particle mass values in dependence on velocity). These equations can be applied to physical reality if Newton's force law is slightly generalized (i.e., the force determines the time change of momentum, not directly that of acceleration). The same holds for Schr\"{o}dinger equation solutions. They represent the superpositions of individual solutions of Hamilton's equations (if these solutions are interpreted on a particle basis). One may ask if it is possible to make use of this generalized classical physics to also describe other phenomena that led previously to the creation of fundamentally different theories of physical reality (see \cref{sec:future_way_ontology}).

It is critical to test all possible consequences of the assumptions of a model of physical reality established on the basis of logical deductions of the observation of matter (i.e., to apply the falsification approach). The invalidity of models that have not withstood falsification must be accepted - and these models must be excluded from further consideration. For physics to move forward, all mistaken assumptions, discussed in this paper and contained in widely used contemporary theoretical descriptions, must be excluded from consideration.

This paper has identified the main conceptual problems in contemporary physical research. Ways to solve them and obtain real scientific knowledge in agreement with reality have been proposed.

\addcontentsline{toc}{section}{References}

%
\end{document}